%% file: sample-acmsmall-conf.tex
\documentclass[acmsmall]{acmart}
\AtBeginDocument{%
  }

\setcopyright{none}
\acmYear{2026}

\usepackage{amsmath,amsfonts}
\usepackage{algorithmic}
\usepackage{graphicx}
\usepackage{textcomp}
\usepackage[ruled,vlined,linesnumbered]{algorithm2e}
\usepackage{multirow}
\usepackage{adjustbox}
\usepackage{pifont}
\usepackage{threeparttable}
\usepackage{booktabs}
\usepackage{bbding}
\usepackage{tablefootnote}
\usepackage{hyperref}
\usepackage{tcolorbox}
\usepackage{enumitem}
\usepackage{url}
\usepackage{soul}
\usepackage{balance}
\usepackage[table]{xcolor}
\usepackage{makecell}
\usepackage{subcaption}
\usepackage{pifont}
\usepackage{ulem}
\usepackage{collcell}
\usepackage{listings}
\usepackage{xcolor}
\usepackage{soul}

\newcommand{\tool}{CATGen}
\newcommand{\tech}{CATGen}

\begin{document}

\title[Context Matters: Improving the Practical Reliability of LLM-Based Unit Test Generation (Experience Paper)]{Context Matters: Improving the Practical Reliability of LLM-Based Unit Test Generation (Experience Paper)}

\author{Junjie Chen}
\orcid{0000-0003-3056-9962}
\affiliation{%
  \institution{Tianjin University}
  \city{Tianjin}
  \country{China}
}
\email{junjiechen@tju.edu.cn}

\author{Ziqi Wang}
\orcid{0009-0000-8630-8724}
\affiliation{%
  \institution{Tianjin University}
  \city{Tianjin}
  \country{China}
}
\email{wangziqi123@tju.edu.cn}

\author{Lin Yang}
\orcid{0000-0002-4475-0925}
\affiliation{%
  \institution{Tianjin University}
  \city{Tianjin}
  \country{China}
}
\email{linyang@tju.edu.cn}

\author{Chen Yang}
\orcid{0000-0003-0759-940X}
\affiliation{%
  \institution{Tianjin University}
  \city{Tianjin}
  \country{China}
}
\email{yangchenyc@tju.edu.cn}

\author{Xiao Chu}
\orcid{0009-0002-9041-7020}
\affiliation{%
  \institution{Huawei Cloud}
  \city{Beijing}
  \country{China}
}
\email{chuxiao1@huawei.com}

\author{Jianyi Zhou}
\orcid{0000-0002-4867-5416}
\affiliation{%
  \institution{Huawei Cloud}
  \city{Beijing}
  \country{China}
}
\email{zhoujianyi2@huawei.com}

\author{Guangtai Liang}
\orcid{0009-0004-2454-1706}
\affiliation{%
  \institution{Huawei Cloud}
  \city{Beijing}
  \country{China}
}
\email{liangguangtai@huawei.com}

\author{Qianxiang Wang}
\orcid{0000-0002-1322-2476}
\affiliation{%
  \institution{Huawei Cloud}
  \city{Beijing}
  \country{China}
}
\email{wangqianxiang@huawei.com}

\author{Dong Wang}
\authornote{Corresponding Author.}
\orcid{0000-0002-2004-0902}
\affiliation{%
  \institution{Tianjin University}
  \city{Tianjin}
  \country{China}
}
\email{dong_w@tju.edu.cn}

\renewcommand{\shortauthors}{J. Chen, Z. Wang, L. Yang et al.}

\input{0_abstract}

\maketitle

\input{1_introduction}
\input{2_related}
\input{3_methodology}
\input{4_industry_evaluation}
\input{5_oss_evaluation}
\input{6_discussion}
\input{7_conclusion}


\normalem
\balance
\bibliographystyle{ACM-Reference-Format}
\bibliography{references}

\end{document}

%% file: 0_abstract.tex
\begin{abstract}

Automated unit test generation has recently benefited from advances in large language models (LLMs), yet our industrial deployments reveal a persistent gap between promising research results and practical usability. In real-world projects with complex frameworks and cross-file dependencies, LLM-generated tests frequently fail to compile, require costly manual repair, or provide unstable coverage improvements. This paper reports our experience in designing, deploying, and evaluating \textbf{CATGen}, a context-aware workflow for LLM-based unit test generation, informed by repeated industrial failures and refinements. Rather than relying on LLMs to infer incomplete project context, we found that compilation robustness critically depends on making project-level dependencies explicit, stabilizing test class scaffolding, and replacing iterative LLM-based repair with lightweight static analysis. These experience-driven insights shaped CATGen’s multi-stage design, which combines structured context retrieval, deterministic test skeleton construction, and program analysis–based post-processing. We evaluate CATGen on real-world complex focal methods from proprietary industrial projects and additionally on the Defects4J benchmark to assess generalizability. Across both settings, CATGen substantially improves compilation success and structural coverage while significantly reducing generation time and token consumption compared to existing LLM-based approaches. Our results demonstrate that reliable LLM-based unit test generation in practice depends less on prompt engineering alone and more on systematic engineering support grounded in real-world development constraints.

\end{abstract}

\ccsdesc[500]{Software and its engineering~Software testing and debugging}
\ccsdesc[300]{Computing methodologies~Natural language generation}
\ccsdesc[300]{Software and its engineering~Software maintenance tools}

\keywords{Unit Test Generation, Large Language Model, Code Context}

%% file: 1_introduction.tex
\section{Introduction}
\label{sec:introduction}

Unit testing is a cornerstone of software quality assurance, detecting bugs early by validating the functionality of each program unit~\cite{zhu1997software,runeson2006survey,almasi2017industrial,yang2025clarifying}.
Manually writing high-quality unit tests for the focal method (a.k.a.\ the method under test) can be tedious and time-consuming~\cite{kumar2016impacts}.
To reduce this effort, numerous automated test generation methods have been developed over the past decades. 
Traditional methods often relied on random-based strategies~\cite{pacheco2007feedback}, constraint-driven techniques~\cite{csallner2008dysy,xiao2013characteristic}, and search-based approaches~\cite{harman2009theoretical, fraser2011evosuite,blasi2022call}.
Deep learning-based methods have emerged~\cite{mastropaolo2021studying,dinella2022toga,nie2023learning}, framing unit test generation as a neural machine translation problem.
Despite their respective strengths, these methods still face limitations in fully understanding the code's intent and generating syntactically correct and effective tests.

Recently, Large Language Model (LLM)-based techniques have gained significant popularity and have shown potential in automated test generation.
Several LLM-based methods have been proposed, including ChatTester~\cite{yuan2024evaluating}, ChatUniTest~\cite{chen2024chatunitest}, HITS~\cite{wang2024hits}, TELPA~\cite{yang2024advancing} and RATester~\cite{yin2025enhancing}. 
These tools empirically demonstrate the strong capability of LLMs to generate high-coverage tests over traditional ones.
For instance, ChatTester, which incorporates an initial test generator and an iterative test refiner, improved the statement coverage from EvoSuite's 68.0\% to 82.3\%, evaluated on their datasets.
HITS is specifically devised to decompose complex focal methods into slices for LLMs, proving its superiority in generating high-coverage unit tests.

However, both prior empirical studies and our industrial experience reveal that compilation failures remain a major barrier to the practical adoption of LLM-based test generation.
Large-scale evaluations~\cite{yang2024empirical} have shown that a significant portion of LLM-generated unit tests fail to compile, regardless of the prompting strategy used~\cite{wang2024hits, schafer2023empirical}.
In practice, such failures severely limit the usefulness of generated tests, as developers must first diagnose and repair compilation errors before any coverage or defect detection benefits can be realized.
Through collaboration with a large global industrial partner and deployment of existing LLM-based test generation on real-world projects (detailed in Section 4), we systematically analyzed recurring failure cases.
Our investigation suggests that insufficient and improperly utilized project-level context is the primary underlying challenge, which manifests in several recurring forms:

\begin{itemize}[leftmargin=15pt,nosep]
    \item \textbf{(I) Context mismatch in industrial project environments.} Real-world software projects involve complex project-level dependencies, including testing frameworks, mocking libraries, and cross-file interactions, that are often poorly captured by existing methods.  
    As a result, LLMs have to infer these dependencies without adequate context, frequently leading to incorrect imports, unresolved symbols, or incompatible framework usage.
    In our deployments, these mismatches frequently caused tests to fail at the first compilation, regardless of test logic quality.
    \item \textbf{(II) Fragility of test scaffolding.} Constructing a correct test class skeleton, including imports, annotations, class declarations, mock definitions, and setup logic, requires strict adherence to framework-specific conventions and precise contextual information~\cite{yang2024empirical}.
    Yet existing approaches typically ask LLMs to generate it from scratch, where even the test logic was plausible, minor skeleton deviations (e.g., missing lifecycle hooks or mismatched mocking annotations) often invalidated the entire test class.
    \item \textbf{(III) Escalating cost of post-generation repair}. To compensate for missing context, existing approaches often rely on iterative LLM-based refinement or post-generation repair~\cite{pan2025aster,alshahwan2024automated,liu2025llm,ni2024casmodatest}. 
    In our industrial deployments, these strategies frequently incurred substantial time and token overhead, while yielding limited improvements in compilation robustness.
\end{itemize}
These limitations are magnified in large-scale industrial scenarios with tight release cycles, where manual correction of compilation errors is costly and undermines the productivity gains promised by automated test generation.
In such settings, effective utilization of project-level context becomes critical for practical adoption at scale.

Motivated by these observations, this paper reports our experience in designing and applying a context-aware LLM-based unit test generation workflow, which we refer to as \textbf{CATGen}.
Our goal is to distill a set of practice-driven design principles informed by repeated failure patterns observed in industrial usage, to advance the practical reliability and applicability of LLM-based unit test generation.
The central insight is that compilation robustness improves when LLMs are systematically supported with explicit, structured context, instead of being asked to infer it implicitly.
Specifically, CATGen is guided by three experience-derived principles centered on context acquisition and utilization via deterministic program analysis:
(I) Instead of leaving LLMs to infer incomplete or missing project-level dependencies, CATGen explicitly retrieves contextual information from project structures and build configurations, ensuring that framework usage and external dependencies are accurately captured.
(II) Rather than forcing LLMs to generate fragile test scaffolding code from scratch, CATGen constructs a valid test class skeleton using context-aware templates and systematic mocking strategies, allowing LLMs to focus exclusively on generating test logic.
(III) To avoid costly iterative refinement, CATGen applies lightweight program analysis–based post-processing rules gained from developer feedback to deterministically repair common compilation errors and enhance test adequacy.

To demonstrate the practical value and generalizability of the proposed approach, we evaluate CATGen in both industrial and open-source environments.
In the \textbf{industrial setting}, we construct a benchmark of eight proprietary projects from our global industrial partner, covering diverse real-world scenarios such as advanced Java features, common design patterns, and framework-intensive systems. Together with partner engineers, we curate 183 focal methods jointly prioritized for maintenance risk (evolving logic) and regression risk (failure-prone interactions); the two notions overlap in practice but are not identical sets.
These methods reflect substantial industrial complexity: 57.38\% involve complex dependencies, and each interacts with 3.05 external files on average. 
All methods originate from actively maintained production code, representing realistic testing demands rather than artificially difficult cases. 
Moreover, because the codebases are proprietary, the benchmark avoids potential LLM data-leakage concerns common in open-source datasets.
We compare CATGen against six representative or state-of-the-art baselines, including one search-based technique and five LLM-based approaches, using compilation success rate, line coverage, branch coverage, and passing rate.
The results show that CATGen consistently achieves substantially higher compilation success, with improvements ranging from 24.72\%-38.05\%, while also significantly improving coverage, yielding 17.27\%–22.17\% gains in line coverage and 15.31\%–18.24\% gains in branch coverage, together with reduced time and token consumption by 51.27\%–69.00\% in time and 66.83\%–83.86\% in token usage.
In the \textbf{open-source setting}, we conduct experiments on Defects4J to examine whether the same design principles generalize beyond proprietary projects. 
CATGen exhibits a consistent performance trend, delivering 10.42\%–14.33\% improvements in compilation success over existing LLM-based approaches, while also achieving 6.11\%–8.39\% higher line coverage and 3.27\%–10.56\% higher branch coverage.
Taken together, these findings highlight how practice-driven system design choices can substantially enhance the practicality of LLM-based unit test generation in industrial settings, while also demonstrating strong potential for adoption in open-source ecosystems.

\textbf{Contributions.} This experience paper makes the following contributions: 
\ding{182} We identify key bottlenecks that limit automated unit test generation in industrial deployments based on real-world experience.
To address them, we curate an industry-grounded benchmark with our partner and design \tech{}, a context-aware workflow that integrates LLM generation with deterministic program analysis.
\ding{183} We conduct an extensive evaluation in both industrial and open-source settings to assess the effectiveness and efficiency of \tech{} against state-of-the-art LLM-based approaches. 
The results show that \tech{} consistently outperforms the baselines across all the evaluated metrics.
\ding{184} We learned that practical LLM-based test generation depends less on prompt engineering alone and more on systematic engineering support: (i) explicitly retrieving project-level context instead of relying on the model to infer dependencies,
    (ii) constructing a stable test-class skeleton rather than generating fragile initialization code from scratch, and
    (iii) applying deterministic, analysis-driven post-processing to avoid costly iterative LLM repair loops.


%% file: 2_related.tex
\section{Related Work}
\label{sec:background}

Automated unit test generation is the process of automatically creating test cases to validate code functionality~\cite{yang2024empirical}. 
Typically, it involves analyzing the source code to identify the focal method and then generating inputs and expected outputs to verify its behavior. 
Over the past decade, numerous automated unit test generation approaches have been proposed to reduce the manual effort required from developers.
Traditional techniques include random-based strategies, search-based approaches, and model checking. 
For example, EvoSuite~\cite{fraser2011evosuite}, one of the most influential test generation techniques, employs an evolutionary algorithm to generate initial test cases and then iteratively refines them using mutation and selection strategies.

To address the readability and maintainability challenges of traditional approaches, DL-based test generation techniques have been developed by framing test generation as a neural machine translation problem~\cite{wang2024software, wang2025survey}.
For instance, \citet{watson2020learning} introduced ATLAS, which leverages neural machine translation to generate meaningful assert statements for test methods automatically. 
Similarly, \citet{tufano2020unit} proposed ATHENATEST, a BART Transformer-based approach that generates unit test cases by learning from real-world focal methods and developer-written test cases.
Although DL–based techniques show promise, their effectiveness is constrained by the reliance on smaller, general-purpose pre-trained models, and they often struggle to ensure syntactic correctness and executability in complex project settings.

LLMs, trained on extensive corpora, have demonstrated impressive capabilities in generating unit tests. 
Several ChatGPT-based techniques have been developed to leverage its powerful language understanding and code generation skills for generating effective unit tests.
For example, \citet{yuan2024evaluating} introduced ChatTester, which iteratively generates unit tests through interactive conversations with ChatGPT.
\citet{haji2024copilot} studied GitHub Copilot for Python test generation in developer workflows, and \citet{lemieux2023codamosa} combined search-based testing with pre-trained LLMs to escape coverage plateaus.
\citet{wang2024hits} produced HITS by decomposing the focal methods into slices and asking the LLM to generate test cases slice by slice.
TELPA~\cite{yang2024advancing}, RATester~\cite{yin2025enhancing}, and WiseUT~\cite{wiseut} further incorporate cross-file context, aiming to improve coverage for complex behaviors.
Beyond coverage-oriented generation, RTED~\cite{rted} improves type error detection through reflective test generation and type constraints, SEGA~\cite{sega} targets business logic bugs by extracting business semantics from requirement documents, and CLAST~\cite{clast} enhances the semantic clarity of generated tests.
On the other hand, \citet{yang2024empirical} conducted the first large-scale study to investigate multiple open-source LLMs, highlighting the impact of prompt design and model choice.

Despite these advances, LLM-based approaches still face significant challenges in practical settings.
In particular, while many methods report improvements in coverage or test quality, they often struggle to generate tests that reliably compile and execute in real-world projects.
Compared to benchmark datasets (e.g., Defects4J), industrial projects involve more complex dependencies, framework constraints, and cross-file interactions, which frequently lead to compilation failures and unstable test behavior, limiting practical usability as developers must manually fix errors before tests can be executed.

Beyond academic benchmarks, recent industrial deployments include Meta's TestGen-LLM~\cite{alshahwan2024automated}, Google's BRT-Agent~\cite{cheng2025agentic}, and Mozilla's BLAST~\cite{kitsios2025automated}.
Those efforts emphasize integrating LLMs into large-scale developer workflows and reporting productivity-oriented outcomes.
They seldom foreground systematic analyses of why generated tests fail to compile in framework-heavy codebases.
Our experience paper complements them by studying recurring compilation failures from proprietary deployments and encoding mitigations in CATGen through structured context retrieval, deterministic scaffolding, and analysis-driven repair aimed at executability as well as coverage.

%% file: 3_methodology.tex
\section{Practice-Driven Workflow for LLM-Based Unit Test Generation}
\label{sec:methodology}

\begin{figure}[t]
    \centering    
    \setlength{\abovecaptionskip}{2pt} 
    \setlength{\belowcaptionskip}{-10pt} 
    \includegraphics[width=\linewidth]{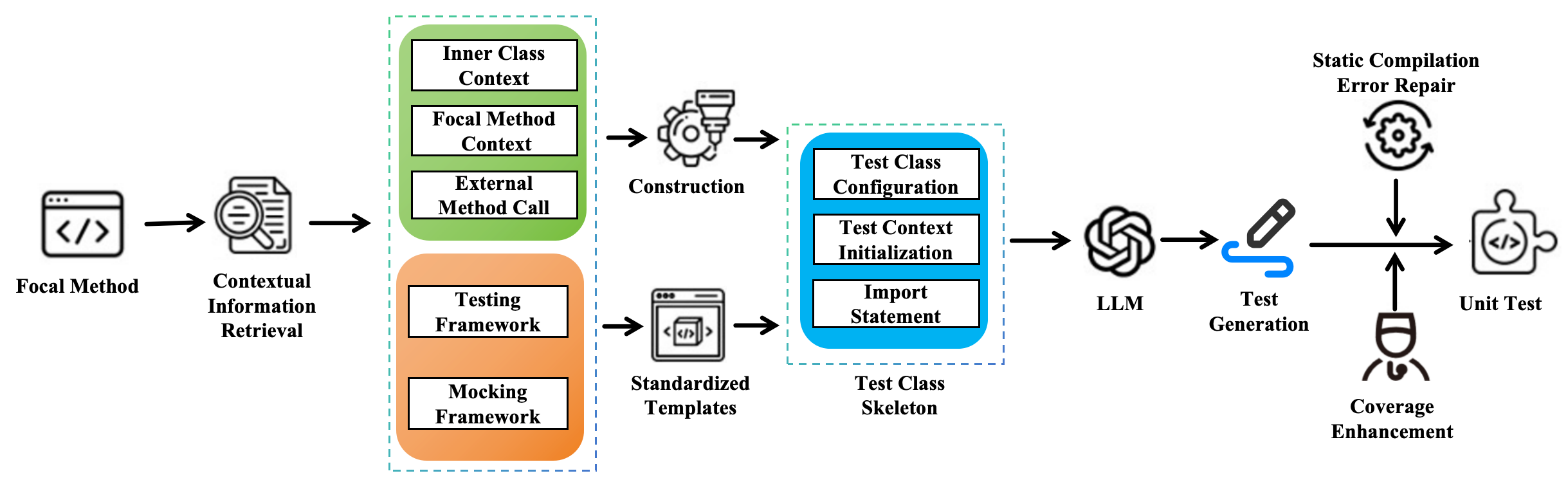}
    \caption{Overview of the proposed \tech{}}
    \label{fig:overview}
\end{figure}

Figure~\ref{fig:overview} depicts the workflow we converged to after pilots on production repositories, iteratively tracing compilation and runtime failures back to missing context or scaffolding.
Once recurring failure classes stabilized under explicit stages, we froze this layout.
In deployment the limiting factor was seldom ``more coverage on one method'' but rather producing tests that compile and execute under frameworks, dependencies, and cross-file interactions---settings where generation without explicit project context frequently fails.
We structure generation into four stages.
Given a focal method, \tech{} (1) retrieves project-level context including build configuration and relevant code elements, so that frameworks, imports, and cross-file calls are not left to model inference (Section~\ref{subsec:3.1}); (2) constructs a test class skeleton tailored to the project’s testing/mocking setup, avoiding fragile scaffolding generation from scratch (Section~\ref{subsec:3.2}); 
(3) generates test methods via skeleton-conditioned completion, where a fixed skeleton anchors the model output and reduces structural drift (Section~\ref{sec:3.3}); and (4) applies program analysis–based post-processing to deterministically repair common compilation issues and improve robustness without relying on costly iterative LLM refinement (Section~\ref{sec:3.4}).

\subsection{Contextual Information Retrieval}
\label{subsec:3.1}

In our early trials, we observed that prompting an LLM with the focal method alone was often insufficient for real-world projects.
The generated tests frequently overlooked framework-specific conventions (e.g., annotations and imports), used incompatible mocking APIs, or invoked external methods without matching signatures. 
These issues typically surfaced as compilation failures and made downstream repair expensive. 
We therefore converged on a lightweight retrieval step that makes project-level dependencies and cross-file behaviors explicit, rather than leaving them to the model’s inference. 
Specifically, \tech{} collects five complementary types of context:
\begin{itemize}[leftmargin=10pt,nosep]
    \item \textbf{Inner Class Context.} Structural and semantic elements of the focal class, including imports, constructors, fields, and public methods. 
    This helps the model build valid instances, manage object state, and trigger intended logic paths.
    
    \item \textbf{Focal Method Context.} Details of the focal method, including its parameter/return types and implementation.
    This helps reduce type mismatches and semantically invalid invocations, which in turn lowers the likelihood of compilation errors and common runtime issues.
    \item \textbf{External Method Call Information.} For methods invoked by the focal method, \tech{} extracts their signatures, types, return values, and method bodies. 
    This provides the behavioral and typing constraints needed for consistent stubbing and for reasoning about reachable paths.
    \item \textbf{Testing Framework.} The project’s testing framework determines the correct annotations and assertion conventions. 
    \tech{} parses build configuration files (e.g., \textit{pom.xml}) to detect framework keywords such as JUnit 5. 
    \item \textbf{Mocking Framework.} The mocking library (e.g., Mockito) is essential for dependency isolation and avoiding mix-and-match errors, also evident by prior studies~\cite{spadini2017mock, zhu2025understanding}.
    Similar to the testing framework, \tech{} infers mocking frameworks by parsing build configuration files.
\end{itemize}

To materialize the five context types, \tech{} first scans project build descriptors for declared testing and mocking artifacts, then parses focal files into ASTs using PSI-compatible structures so signatures, fields, and intra-class members remain faithful to the workspace view.
For outgoing calls from the focal method, \tech{} traverses call expressions to resolve targets where symbol resolution succeeds; it records visibility and modifiers because they constrain whether a dependency must be mocked, injected, or exercised via reflection later.
Unresolved or library-only symbols are retained conservatively so that skeleton imports and mock hooks remain aligned with what the compiler must see.
Taken together, this retrieved context grounds skeleton construction and post-processing; extraction uses lightweight static analysis.

\subsection{Context-Aware Test Class Skeleton Construction}
\label{subsec:3.2}

\begin{figure}[tbp]
    \centering
    \setlength{\abovecaptionskip}{-2pt}
    \setlength{\belowcaptionskip}{-3pt}

    \begin{subfigure}{0.85\linewidth}
        \centering
        \setlength{\abovecaptionskip}{0pt}
        \includegraphics[width=0.95\linewidth]{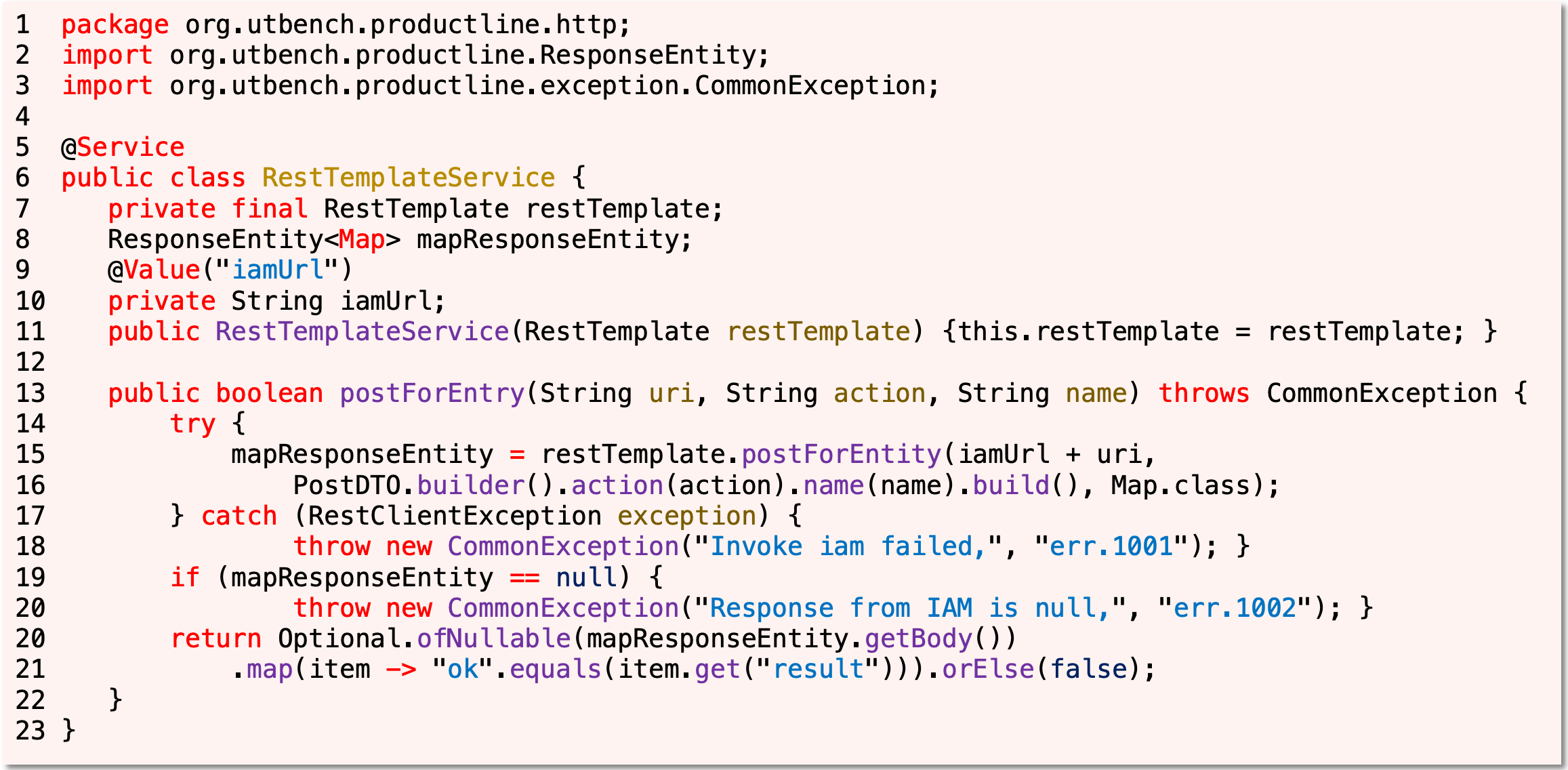}
        \caption{Focal method}
        \label{fig:focal}
    \end{subfigure}
    \vspace{0.5em}
    \begin{subfigure}{0.8\linewidth}
        \centering
        \setlength{\abovecaptionskip}{0pt}
        \includegraphics[width=\linewidth]{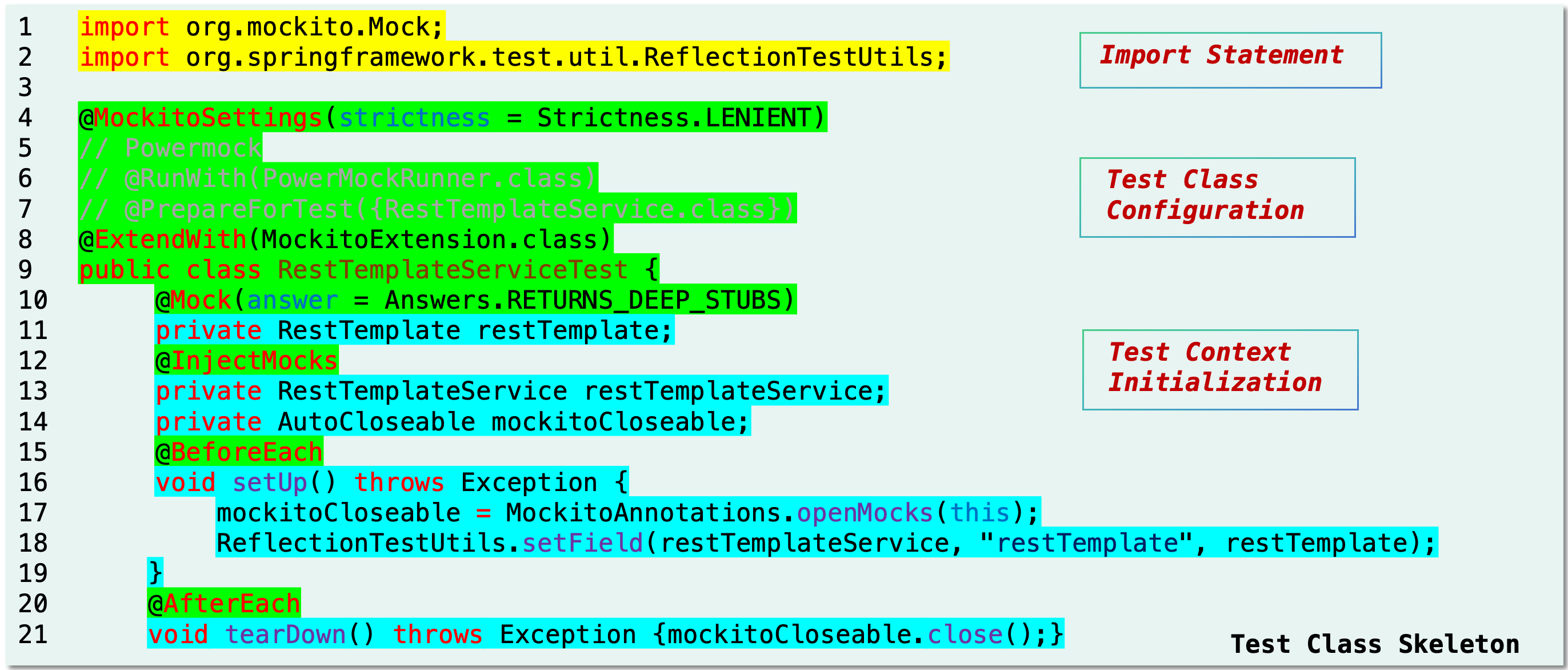}
        \caption{Test class skeleton}
        \label{fig:skeleton}
    \end{subfigure}
    \vspace{0.5em}
    \caption{The illustrative focal method, along with its corresponding test class skeleton generated by \tech{}}
    \label{fig:combined}
    \vspace{-5pt}
\end{figure}

In industrial projects, we found that the test class skeleton is often a decisive factor for usability, as it establishes the execution environment for all test methods.
Even minor omissions, such as missing imports, incorrect framework annotations, or incomplete initialization, can invalidate the entire test class, regardless of whether the generated assertions themselves are reasonable.
During our early attempts, when LLMs were asked to generate the skeleton from scratch, they frequently hallucinated framework-specific patterns or dependency configurations.
Based on these observations, we treat the skeleton as a construction task rather than a free-form generation problem, and instead build it deterministically using the retrieved project context.

CATGen constructs a reliable skeleton via a framework-to-template mapping. Each template encodes the framework-required annotations, lifecycle methods, and configuration patterns, curated from official documentation, so that the skeleton consistently conforms to framework conventions. 
As illustrated in Figure~\ref{fig:focal}, we use the focal method \textit{RestTemplateService.postForEntry(String, String, String)} from our industrial benchmark as a running example. 
This method interacts with external dependencies and includes error-handling logic, making it representative of cases where scaffolding mistakes are common. Figure~\ref{fig:skeleton} shows the corresponding skeleton constructed by \tech{}, which consists of three parts:

\noindent \textbf{Test Class Configuration.}
Skeleton templates abstract framework-level structural patterns (imports, lifecycle hooks, and mocking APIs) instantiated per detected stack.
The configuration establishes the structural and annotative foundation of the test class and ensures compatibility with the selected testing and mocking frameworks. 
\tech{} derives the test class name by appending \textit{“Test”} to the focal class name, and sets visibility according to framework conventions (e.g., \textit{public} under JUnit~5) to support test discovery. It then adds class-level annotations based on detected dependencies: for Mockito-based tests, \textit{@ExtendWith(MockitoExtension.class)} enables \textit{@Mock} and \textit{@InjectMocks}; for PowerMock-based tests, when static mocking is required, declaring classes are added to \textit{@PrepareForTest}. At the field level, Mockito templates introduce \textit{@Mock} for dependencies and \textit{@InjectMocks} for wiring them into the focal object. Finally, lifecycle hooks (e.g., \textit{@BeforeEach}, \textit{@AfterEach}) are added to standardize setup/teardown for isolation and repeatability. We provide detailed construction rules in our replication package. The green region in Figure~\ref{fig:skeleton} illustrates the resulting configuration.

\noindent \textbf{Test Context Initialization.}
Beyond configuration, the skeleton must correctly initialize the focal object and its dependencies so that generated test methods can execute reliably. 
\tech{} constructs the test context by mapping focal-class member variables to test-class fields and ensuring constructors are invoked or configured appropriately. In the example (blue region in Figure~\ref{fig:skeleton}), using JUnit~5 and Mockito, \tech{} declares the focal class instance as a field annotated with \textit{@InjectMocks} (lines 13–14), and introduces \textit{@Mock}-annotated fields for dependencies (line 11). A \textit{@BeforeEach} method (lines 16–19) initializes mocks and sets non-mockable or configuration-dependent fields via reflection. An \textit{@AfterEach} method (line 21) is also generated to keep the skeleton complete and facilitate cleanup.

\noindent \textbf{Import Statements.}
Scaffold imports are assembled deterministically from three sources: \emph{(i)} framework imports implied by the detected testing/mocking stack; \emph{(ii)} standard assertion and utility imports for the chosen framework; and \emph{(iii)} dependency imports resolved from focal-method and focal-class types via static analysis and classpath information, rather than being inferred by the LLM (Figure~\ref{fig:skeleton}, yellow).
When resolution is ambiguous we conservatively prefer project-local and framework-consistent symbols; imports for types introduced only in LLM-completed bodies are supplemented or corrected in Section~\ref{sec:3.4}.

\subsection{Skeleton-Conditioned Completion for Test Generation}
\label{sec:3.3}

A common LLM-based test generation workflow follows a dialog-style paradigm: users provide a natural language query, and the model outputs candidate tests~\cite{wang2024hits, yuan2024evaluating, chen2024chatunitest}.
While straightforward, our experience suggests that this interaction pattern is fragile in practice.
In a pilot study, we observed that even when a pre-constructed test skeleton is provided, LLMs
(including advanced models such as GPT-4) often drift from the given structure, producing
incomplete or invalid test cases. This drift frequently manifests as missing required boilerplate,
redefining class-level elements inconsistently, or returning partial snippets that are hard to compile
and integrate. 
To mitigate these issues, we reformulate test generation as a skeleton-conditioned completion task rather than generating an entire test class from scratch. 
Concretely, we decompose the input into two complementary components:

\noindent \textbf{Contextual Prompt.}
The prompt provides structured guidance for generating test methods. 
It includes class-level information (class name, constructors, member fields) to support instantiation and state management, followed by the focal method implementation and related methods to enable behavior reasoning. 
It also specifies the detected testing and mocking frameworks and instructs consistent handling of static and non-static dependencies. 
Finally, the prompt requires the model to analyze the branching structure of the focal method before writing tests to encourage systematic exploration of control flow and improve branch coverage.

\noindent \textbf{Prefilled Content.}
We prefill the response with the test class skeleton constructed in Section~\ref{subsec:3.2}. 
The model is then asked to complete the remaining parts under this fixed scaffold: it writes test methods, configures mock behaviors, and inserts assertions while preserving the skeleton’s structure. 
In our experience, anchoring generation in this way reduces syntactic errors and structural drift, and increases the likelihood that the output is directly compilable.
The full prompt is provided in our replication package.

\begin{figure}[t]
    \centering    
    \setlength{\abovecaptionskip}{2pt} 
    \setlength{\belowcaptionskip}{-10pt} 
    \includegraphics[width=0.8\linewidth]{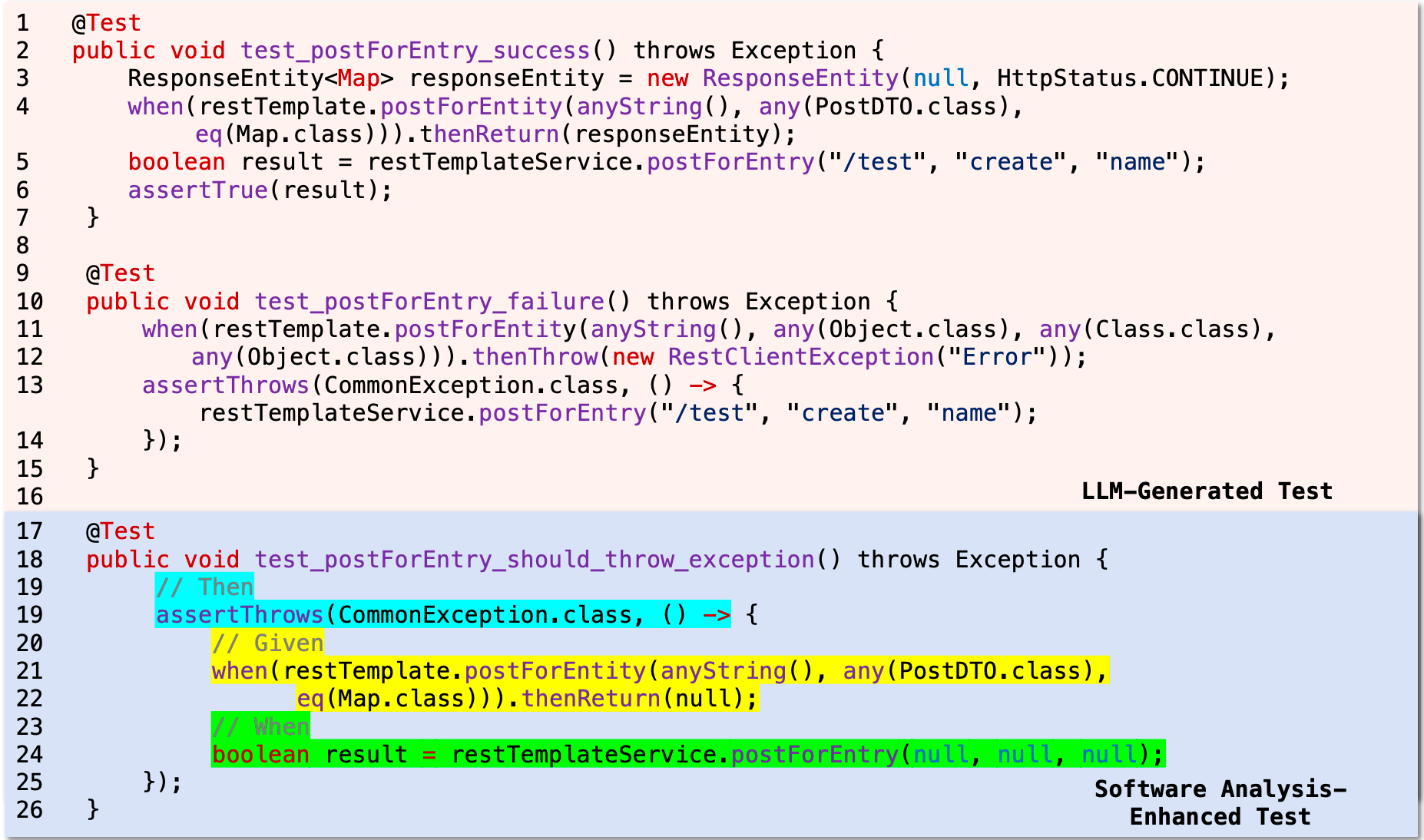}
    \caption{Test Class Generated by \tech{}}
    \label{fig:test}
    \vspace{-5pt}
\end{figure}

\subsection{Program Analysis-Based Post-Processing}
\label{sec:3.4}

LLM-generated tests often require post-processing before they become usable, and a common strategy is to iteratively re-prompt the model with compiler feedback~\cite{chen2024chatunitest}.
In our early deployments, we adopted the same feedback-driven repair loop and ran multiple LLM repair rounds to fix compilation errors.
However, this approach proved brittle and cost-unstable in practice: when hallucinations persisted, or project-level dependencies were only partially captured, similar errors repeatedly resurfaced across iterations, while token usage and latency increased rapidly as the loop continued.
To make repair predictable and cost-stable, \tech{} instead adopts deterministic, lightweight post-processing grounded in program analysis and the retrieved project context.
The process consists of two stages:

\noindent \textbf{Static Compilation Error Repair.}
We observe three dominant categories of compilation failures in the LLM-completed test bodies: (I)~unresolved dependencies, including missing imports, incorrect package qualifiers, or ambiguous simple names; (II)~references to undefined types, methods, or variables; and (III)~incorrect initialization or framework usage (e.g., Mockito setup or constructor misuse).
\tech{} implements a lightweight repair pipeline on top of the project context from Section~\ref{subsec:3.1} and compiler-visible facts from AST inspection, so that repairs are grounded in the same dependency and signature information as skeleton construction, rather than in additional LLM rounds.

Following empirical observations of industrial developer repair workflows and insights from prior work~\cite{yuan2024evaluating}, we implement a suite of targeted strategies:
(1)~\textit{Package Declaration Completion}, which aligns the test class package with the focal method's source location;
(2)~\textit{Import Statement Supplementation}, which resolves simple names and static members against the focal compilation unit, neighboring files in the same module, and the classpath slice from context retrieval, and inserts or adjusts \texttt{import} declarations when resolution is unique;
(3)~\textit{Class Annotation Rectification}, which corrects JUnit and Mockito annotation configurations;
(4)~\textit{Invalid Reference Resolution}, which removes or rewrites references that cannot be bound to symbols visible through static analysis;
(5)~\textit{Private Member Access Adaptation}, which introduces reflection-based access when private members must be exercised;
(6)~\textit{Method Signature Alignment}, which reconciles method names and parameter or return types with their actual signatures;
(7)~\textit{Exception Specification Enhancement}, which adds explicit exception declarations or assertions where required for compilation or consistent exception handling; and
(8)~\textit{Fallback Assertion Mechanism}, which introduces default assertions so that test methods retain minimal behavioral checks.

All eight strategies share the same backbone: each rule is driven by compiler diagnostics and lightweight AST inspection over the project context from Section~\ref{subsec:3.1}, fires in a fixed precedence order, and never re-invokes the LLM, which keeps repeated runs deterministic and cost-bounded.
Building on this backbone, the three failure categories are dispatched to complementary rules.
For~(I), import supplementation extends the scaffold imports of Section~\ref{subsec:3.2} to types that surface only in LLM-completed bodies; when a binding is ambiguous it conservatively prefers project-local, framework-consistent symbols and otherwise hands the case off to invalid-reference resolution rather than inventing packages.
Building on this, category~(II) is resolved by the same invalid-reference resolution working together with method signature alignment, which together reconcile unbound calls and mismatched names or parameter/return types against the signatures recovered from the focal class and its dependencies.
Finally, category~(III) is addressed by class annotation rectification, private member access adaptation, exception specification enhancement, and the fallback assertion mechanism, which repair JUnit/Mockito scaffolding, private-access patterns, and \texttt{throws}/assertion structure so that the merged class compiles and exhibits minimally consistent behavior.
Overall, this stage is deliberately limited to restoring compilability and structural consistency; broader test adequacy is evaluated via coverage and mutation analyses later in the paper, and the complete rule ordering with worked examples is provided in the replication package.

\noindent \textbf{Program Analysis-Based Coverage Enhancement.}
Even when tests compile, they may miss corner cases.
To complement this gap without additional LLM calls, \tech{} uses static analysis to identify critical decision points such as null checks, empty-string validations, and explicit exception throws.
For each uncovered condition, CATGen synthesizes an additional
test following the Given--When--Then paradigm~\cite{yuan2024evaluating,wang2024hits}:
in the Given phase, it analyzes exception-handling
structures and external dependencies and simulates them via mocking; 
in the When phase, it injects
boundary/invalid inputs (e.g., \textit{null} or empty strings) to trigger the targeted path; and in the
Then phase, it generates value-based and exception-aware assertions (e.g., \textit{assertTrue},
\textit{assertThrows}) to validate the expected behavior. 
Figure~\ref{fig:test} illustrates an example where
LLM-generated tests miss a null-input scenario, and the enhancement module automatically
adds a targeted test to exercise and validate the corresponding exception handling. 
Finally, \tech{} merges analysis-driven tests into the LLM-completed class.
A deterministic consolidation step resolves colliding \texttt{@Test} names (suffixing enhancement-only methods when needed), removes near-duplicate bodies, and orders methods for readability; static repair then reconciles imports and signatures on the merged class.

%% file: 4_industry_evaluation.tex
\section{Industrial Evaluation}
\label{sec:industrial}

\subsection{Experimental Design}

\textbf{Industrial Benchmark.}
\input{Tables/dataset}
To ensure that our evaluation reflects the challenges we encountered in production settings, we construct the benchmark in close collaboration with our industrial partner, capturing the failure modes observed when deploying LLM-generated tests in real repositories.
In these codebases, a focal method is rarely self-contained. Its behavior is often shaped by dependency injection and configuration wiring, by contracts that are distributed across files and modules, and by framework managed execution paths where correctness depends on conventions and implicit resources. 
We also saw many methods with non-linear control and data flow, such as nested conditionals, early returns, and chained transformations, where meaningful branches are difficult to reach with simple happy path inputs. When this surrounding context is missing or only partially inferred, the generated tests tend to either fail to compile, set up mocks inconsistently, or run but exercise little of the intended behavior. These observations are why we treat compilation success and cost stability as first order concerns in this benchmark, not secondary metrics.

To capture these practical realities, we curate the benchmark together with engineers from our industrial partner using actively maintained and production-deployed internal Java projects.
The projects span multiple architectural layers, including core service logic, data access code, configuration and wiring components, web controllers, shared utilities, and microservice infrastructure. 
We use project-level proportional stratified sampling to reflect both scale and diversity, and we intentionally bias selection toward difficult methods with deep call chains, cross-module interactions, and framework-managed resources. 

The benchmark totals 183 focal methods (Table~\ref{tab:dataset-statistics}): \textbf{57.38\%} have complex dependencies and \textbf{3.05} dependent files on average.
We count distinct external classes and methods invoked by the focal method as a coupling-oriented proxy aligned with coupling between objects (CBO)~\cite{chidamber1994metrics}, and treat $\geq 3$ such dependencies as \textit{complex dependencies}, reflecting dependency-heavy cases where our deployments saw frequent compilation failures.
The \textit{Microservice Integration Infrastructure} scenario is the hardest subset (\textbf{100\%}; \textbf{5.44} files on average).

\noindent \textbf{Compared Techniques.}
To demonstrate the effectiveness of the proposed workflow, we comprehensively adopt six representative or state-of-the-art test generation techniques as baselines:

\begin{itemize}[leftmargin=10pt,nosep]
    \item \textbf{EvoSuite}~\cite{fraser2011evosuite}. A traditional search-based tool that generates JUnit tests via evolutionary algorithms; it iteratively executes and mutates candidate tests to maximize structural coverage (e.g., line and branch) without using LLMs.
    \item \textbf{ChatTester}~\cite{yuan2024evaluating}. It includes an initial test generator and an iterative test refiner. The initial test generator first leverages an LLM to understand the focal method and generate a test, and the refiner then iteratively fixes the compilation errors.
    \item \textbf{ChatUniTest}~\cite{chen2024chatunitest}. It generates unit tests using an LLM by leveraging the focal method and context extracted through predefined rules as inputs. For tests that fail execution, it utilizes JVM error reports to guide the LLM in correcting them.
    \item \textbf{HITS}~\cite{wang2024hits}. It first decomposes the focal method into slices and creates
    unit tests for each slice, designed to cover all lines and branches.
    These unit tests collectively form the initial test suite. 
    For the non-executable test suite, HITS includes a fixer to repair them. 
    \item \textbf{TELPA}~\cite{yang2024advancing}. It enhances LLM-based test generation with program analysis to target uncovered branches. TELPA extracts object construction sequences and branch-relevant dependencies, and iteratively guides the LLM with coverage feedback.
    \item \textbf{RATester}~\cite{yin2025enhancing}. It improves repository awareness by querying a language server for symbol definitions and usages. 
    The retrieved global context is injected into the prompt to reduce hallucinations and signature mismatches.
\end{itemize}

For all LLM-based baselines, we use their publicly available open-source implementations. Because some original studies instantiated their methods with proprietary models, we standardize all LLM calls to the same set of open-source models to ensure a fair comparison. This standardization is applied consistently wherever an LLM is involved, including both unit test generation and subsequent repair steps.
RATester was originally implemented for Go. To maintain a consistent Java-based evaluation environment, we follow the authors’ Java adaptation and re-implement RATester accordingly, without modifying its underlying algorithmic design.

\noindent \textbf{Evaluation Metrics.}
We evaluated the performance of \tech{} and baselines using four commonly adopted metrics: \textbf{Compilation Success Rate (CSR)}, \textbf{Line Coverage (CovL)}, \textbf{Branch Coverage (CovB)}, and \textbf{Pass Rate (PR)}.
CSR is determined by the ratio of test methods that compile successfully to the total number of test methods generated for all focal methods.
CovL is calculated as the number of covered lines divided by the total number of executable lines in the source code, while CovB is defined as the number of covered branches divided by the total number of branches in the control flow graph.
PR adapts this metric by using the count of test methods that pass execution as the numerator.
For a fair comparison, \tech{} and all baselines are required to generate a single test class, but multiple test methods are allowed within that class. 
For each test class that compiles and executes successfully, we employ the Jacoco tool\footnote{https://github.com/jacoco/jacoco} to measure its line and branch coverage. 

\noindent \textbf{Implementation and Environment.}
\tech{} is implemented in Java. For program structure analysis, it uses \texttt{PsiMethod} from IntelliJ’s PSI to parse ASTs, enabling precise extraction of method signatures, control-flow dependencies, and inter-file relationships.
The workflow depends only on AST-level facts; any conforming extractor supplying the same structural signals can substitute.
The context-aware skeleton construction component employs standardized templates for JUnit~4~\cite{junit4}/5~\cite{junit5}, Spock~\cite{spock}, and Spring Boot Test~\cite{spring-boot-test}, with Mockito~\cite{mockito} and PowerMock~\cite{powermock} for mocking.

From an engineering perspective, our goal is to evaluate \tech{} under practically deployable open-source LLMs rather than exhaustively benchmark all available models. We therefore select model families that (i) are publicly accessible and reproducible, (ii) provide instruction-tuned checkpoints suitable for interactive code generation, and (iii) cover both general-purpose and code-specialized variants across multiple parameter scales.
We evaluate models from the \textbf{Llama}, \textbf{DeepSeek}, and \textbf{Qwen} families, spanning both general and code-specialized models across multiple scales:
\textbf{CodeLlama-7B (CL-7B)}~\cite{codellama-7b-hf}, \textbf{Llama3.1-8B (Lla-8B)}~\cite{llama3-8b-hf},
\textbf{DeepSeekCoder-6.7B-Instruct (DC-7B)}~\cite{deepseekcoder-6.7b-hf}, \textbf{DeepSeekCoder-33B-Instruct (DC-33B)}~\cite{deepseekcoder-33b-hf},
\textbf{DeepSeek-R1-Distill-Llama-8B (DRL-8B)}~\cite{deepseekr1-llama-8b-hf}, \textbf{DeepSeek-R1-Distill-Qwen-32B (DRQ-32B)}~\cite{deepseekr1-qwen-32b-hf},
\textbf{Qwen2.5-Coder-7B-Instruct (QC-7B)}~\cite{qwen2.5-coder-7b-hf}, \textbf{Qwen2.5-Coder-32B-Instruct (QC-32B)}~\cite{qwen2.5-coder-32b-hf}, and \textbf{Qwen2.5-32B (Q-32B)}~\cite{qwen2.5-32b-hf}.

\tech{} is deployed in a hybrid execution environment consisting of 4 NVIDIA A100-PCIE-40GB GPUs (driver version 535.161.08, CUDA 12.2) and an Intel Xeon Gold 6330 CPU.
The software stack includes JDK 11.0.8, Gradle 7.4, and Maven 3.3.9 for managing Java modules, with JaCoCo 0.8.11 integrated for code coverage analysis.
Python 3.10.12 is used for LLM-related tasks, and LLM inference is performed using vLLM 0.8.4~\cite{vllm} for efficient and scalable deployment.
All experiments use a zero-temperature setting to ensure deterministic outputs. For the EvoSuite baseline, we allocate a search budget of 300\,s per focal method~\cite{fraser2011evosuite}.

\subsection{Evaluation Results}


\noindent \textbf{I. Effectiveness of \tech{} in industrial setting.}
To mitigate stochasticity in LLM operations and guarantee result reliability, we independently ran all LLM-based test generation processes five times with the same configuration and used the per-method average as the final result for each metric.
To assess the robustness of the observed improvements, we use Wilcoxon signed-rank tests~\cite{woolson2007wilcoxon} on paired per-method outcomes after averaging the five repetitions.

\input{Tables/rq1}

\textbf{Finding I: CATGen substantially improves executability, while also delivering higher structural coverage than existing methods in the industrial scenarios.}
From our experience, a critical requirement in industrial projects is whether generated tests can reliably compile and execute under project-specific configurations, as this directly determines how much behavior the tests can meaningfully exercise and validate.
As shown in Table~\ref{tab:rq1}, for compilation success rate, CATGen reaches 91.83\%, while EvoSuite reaches 75.80\%, and the LLM-based baselines achieve 55.88\% for ChatTester, 51.70\% for ChatUniTest, 51.02\% for HITS, 64.11\% for TELPA, and 67.11\% for RATester. 
EvoSuite occasionally exhausts its budget without emitting a test on highly coupled focal methods, which we count as compilation failures and which partly explains its lower compilation success rate~\cite{fraser2011evosuite}.
For coverage, CATGen reaches 70.10\% on line coverage and 63.92\% on branch coverage. RATester achieves 48.43\% and 46.33\%, TELPA achieves 45.99\% and 44.32\%, and EvoSuite achieves 39.55\% and 34.56\%. ChatTester achieves 30.76\% and 29.75\%, ChatUniTest achieves 29.77\% and 26.74\%, and HITS achieves 36.10\% and 36.33\%. 
For pass rate, CATGen reaches 54.63\%, while EvoSuite reaches 44.21\%, RATester reaches 41.71\%, TELPA reaches 39.91\%, ChatTester reaches 35.85\%, HITS reaches 32.93\%, and ChatUniTest reaches 30.24\%. 
Notably, among the LLM-based baselines, TELPA and RATester tend to achieve stronger overall results, which aligns with their designs that explicitly incorporate broader project context and cross-file information when generating tests. 
The statistical testing results confirm that \tech{} consistently outperforms all baselines, with \textit{p-values} $< 0.005$ for CSR, CovL, CovB, and PR in all pairwise comparisons.
We report unadjusted $p$-values for pairwise comparisons; given the consistently small $p$-values observed across all subjects, the statistical significance of our results remains robust.
As an illustrative pairwise contrast for Wilcoxon reporting, for CovL under QC-32B versus RATester, we obtain $n{=}183$, $W{=}1148$, $p{=}0.0026$, and $r{=}0.339$ (medium effect).

\textbf{Finding II: CATGen ensures robust performance across LLMs of varying scales and architectures, with larger models potentially yielding better results.}
As shown in Table \ref{tab:rq1}, the underlying LLM choice significantly impacts all test generation methods: larger models (e.g., 32B–33B) consistently outperform smaller ones like Lla-8B. Among model families, Qwen variants lead overall: QC-32B achieves the highest coverage (CovL: 75.79\%, CovB: 73.73\%), while DRQ-32B attains the best compilation success rate (96.29\%) and pass rate (62.36\%). 
DeepSeek models follow closely, whereas Llama-family models, especially Lla-8B, perform poorly under baselines.
On the other hand, results show that code-specialized variants tend to outperform their general-purpose counterparts at the same scale. 
For instance, QC-32B outperforms Q-32B, and models enhanced via reinforcement learning-based knowledge distillation (e.g., DRQ-32B) further exceed their general counterparts. 
These findings underscore the significance of domain-specific pretraining and distillation-based capability transfer for generating high-quality tests.
Importantly, \tech{} maintains consistent superiority across all LLM configurations. This robustness makes it particularly valuable for cost-sensitive or resource-constrained scenarios. Even with a 7B-scale model like DC-7B, \tech{} outperforms all baselines across all model sizes, offering a compelling balance between efficiency and effectiveness.

\noindent \textbf{II. Contributions of Components in \tech{}.}
We conducted an ablation study to examine the contribution of each core component (i.e., context-aware test class skeleton construction, static compilation error repair, and program analysis-based test enhancement) to the overall effectiveness of \tech{}, with the program analysis-based post-processing module evaluated via its two subcomponents separately. Accordingly, we constructed four variants of \tech{}:
\begin{itemize}[leftmargin=10pt,nosep]
    \item \textbf{``w/o skeleton''} (without the context-aware test class skeleton construction): This variant omits the rule-based test class skeleton from the prompt, instructing the LLM to generate test code from scratch without any structural guidance. 
    \item \textbf{``w/o repair''} (without the static compilation error repair): This variant disables the static repair stage and applies only the program analysis-based test enhancement to the raw test class generated by the LLM.
    \item \textbf{``w/o enhancement''} (without the program analysis-based test coverage enhancement): This variant disables the test coverage enhancement component, retaining test class skeleton construction and static repair, but excluding any analysis-driven enhancements.
    \item \textbf{``w/o all''}: This variant removes all three core components, relying solely on standard prompt-based generation without any additional post-processing or enhancement.
\end{itemize}
Based on the RQ1 results, we select QC-32B as the representative LLM for our ablation study, given its superior line coverage and relatively high compilation success rate, which demonstrate its overall effectiveness.
To mitigate random variation, we repeat each experiment five times and report the average results across all evaluation metrics.

\input{Tables/rq2}

\textbf{Finding III: Each core component of CATGen serves a unique and critical function in achieving the framework’s overall performance.}
Table~\ref{tab:rq2} presents the effectiveness comparison between \tech{} and its four ablated variants.
First of all, removing the context-aware test class skeleton construction (``w/o skeleton'') led to a substantial performance drop: line coverage (CovL) decreased by 16.37\%, branch coverage (CovB) by 10.28\%, compilation success rate (CSR) by 3.04\%, and passing rate (PR) by 8.86\%, compared to the full version. 
This highlights the critical role of the structured test class skeleton in guiding the LLM. Without it, the model is more prone to hallucination, failing to correctly construct class declarations, mock annotations, or instantiate focal classes, resulting in incomplete or invalid test classes. 
Interestingly, while coverage and correctness decline, the relatively stable CSR indicates that program analysis-based post-processing remains effective in preserving compilability, even when initialization is missing.

Second, removing the static compilation error repair stage (``w/o repair'') caused a moderate drop in coverage (CovL: –8.95\%, CovB: –6.22\%) but a much steeper decline in CSR (–16.79\%) and PR (–8.76\%). 
This discrepancy underscores the importance of static repair in addressing syntactic and structural errors that LLMs frequently produce. As this variant retains the other components, including test enhancement, the observed decrease primarily stems from compilability issues rather than changes in the generation workflow.

Third, disabling the program analysis-based enhancement (``w/o enhancement'') resulted in a 10.51\% and 9.51\% drop in CovL and CovB, respectively—comparable to the loss observed when test class initialization is removed. However, CSR and PR remained largely stable (–1.96\% and –0.13\%), suggesting that the generated tests, although less diverse, remain syntactically and semantically valid. This validates that the enhancement module is instrumental in enriching test diversity and uncovering edge-case behaviors.

Finally, the full ablation scenario (``w/o all'')—which removes all three components—resulted in the most severe degradation. CSR and CovL dropped by 38.88\% and 30.06\%, respectively, roughly equating to the cumulative impact of removing individual components. 
These results emphasize the complementary roles of \tech{}'s core modules: the test class skeleton ensures syntactic structure, static repair improves correctness and compilability, and program analysis-based enhancements enhance coverage through targeted test generation. 
Collectively, they form a cohesive, multi-stage framework that substantially improves both the quality and reliability of LLM-generated unit tests.

\noindent \textbf{III. Efficiency of \tech{}{}.}
We evaluate two key metrics to assess computational efficiency and resource utilization: execution time (\textbf{Time}) and token consumption (\textbf{Tokens}). 
These metrics are measured across the generation phase (\textbf{Gen}), the post-processing phase (\textbf{Post}), and their aggregate total (\textbf{Total}).
Following the setup in RQ2, we select QC-32B for validation, as it achieved the best overall performance among the evaluated LLMs. Each experiment is repeated five times to mitigate variability, and the average results are reported.

\input{Tables/rq3}

\noindent \textbf{Finding IV: CATGen achieves substantial efficiency gains over all baselines in both execution time and token consumption, across both the generation and post-processing phases.}
Table~\ref{tab:rq3} summarizes the results for efficiency-related metrics.
In terms of time efficiency, \tech{} completes the full pipeline in 1,836 seconds (the sum of generation and post-processing time), outperforming ChatUniTest (3,768s, –51.27\%), ChatTester (4,471s, –58.93\%), and HITS (5,924s, –69.00\%). It also substantially outperforms TELPA (4,386s, –58.13\%) and RATester (5,217s, –64.80\%), and is markedly faster than the traditional baseline EvoSuite (10,980s, –83.28\%).
These improvements stem from two key design choices: (1) a single-round LLM generation strategy (1,759s), which avoids the costly iterative refinement loops used by baselines; and (2) a program analysis-based post-processing module that requires only 77 seconds, more than one order of magnitude faster than the LLM-driven repair stages in ChatTester (1,666s), ChatUniTest (2,166s), and HITS (2,951s). In contrast, TELPA and RATester rely on multi-round LLM generation without an explicit repair phase, leading to high cumulative latency due to repeated model invocations, while EvoSuite spends most of its time on evolutionary search and repeated test executions rather than model inference.

\tech{} consumes only 203k tokens in total, with zero tokens used in post-processing. This represents a 77.22\% reduction compared to ChatTester (891k) and an 83.86\% reduction compared to HITS (1,258k). Even ChatUniTest, despite its relatively fast generation time, incurs 1,091k tokens due to heavy reliance on LLM-based post-processing. Compared with TELPA (612k) and RATester (735k), \tech{} also achieves substantial token savings, as their multi-round generation strategy requires significantly more LLM interactions.
Collectively, these results demonstrate that \tech{}’s decoupled architecture, which separates test generation from LLM-dependent post-processing, enables deterministic, scalable, and resource-efficient unit test synthesis. 
This design is particularly well-suited for industrial deployment, where latency and compute cost are critical constraints.

%% file: Tables/dataset.tex
\begin{table}[t]
\centering
\caption{Statistics of the evaluation dataset.}
\label{tab:dataset-statistics}
\vspace{-10pt}
\footnotesize
\setlength{\tabcolsep}{1.5pt}        
\renewcommand{\arraystretch}{1.0}  
\begin{tabular}{@{}lccc@{}}
\toprule
Scenario & \# Focal Methods & \% Complex Dep. & \textit{Avg.} Dep. Files \\
\midrule
Service Orchestration Rules & 43 & 34.88 & 1.05 \\
Data Access Persistence & 38 & 78.95 & 4.13 \\
Configuration Dependency Wiring & 12 & 58.33 & 3.42 \\
Framework Entry & 42 & 52.38 & 3.36 \\
Data Transformation Mapping & 16 & 50.00 & 3.81 \\
Collection Stream Processing & 11 & 54.55 & 2.45 \\
Shared Utilities Libraries & 12 & 66.67 & 3.17 \\
Microservice Integration Infrastructure & 9 & 100.00 & 5.44 \\
\midrule
Total                       & 183 & 57.38  & \bfseries 3.05 \\
\bottomrule
\end{tabular}
\vspace{-8pt}
\end{table}

%% file: Tables/rq1.tex
\begin{table}[ht]
\centering
\renewcommand{\arraystretch}{1.2}
\setlength{\tabcolsep}{2pt}

\caption{Effectiveness comparison among different models and different methods.}
\label{tab:rq1}
\vspace{-10pt}
\footnotesize
\resizebox{\textwidth}{!}{
\begin{tabular}{@{}l|l|c*{9}{c}@{}}
\midrule
\textbf{Metric} & \textbf{Method} & \textbf{Avg} & \textbf{CL-7B} & \textbf{Lla-8B} & \textbf{DRL-8B} & \textbf{QC-7B} & \textbf{QC-32B} & \textbf{Q-32B} & \textbf{DRQ-32B} & \textbf{DC-7B} & \textbf{DC-33B} \\
\midrule

\multirow{7}{*}{\textbf{CSR}}
& \textbf{EvoSuite}    & \underline{75.80\%} & N/A & N/A & N/A & N/A & N/A & N/A & N/A & N/A & N/A \\
& \textbf{ChatTester}  & 55.88\% & 29.17\% & 4.60\% & 32.62\% & 66.92\% & 73.03\% & 62.37\% & 75.00\% & 79.23\% & 80.00\% \\
& \textbf{ChatUniTest} & 51.70\% & 32.92\% & 5.41\% & 33.50\% & 63.53\% & 71.06\% & 46.09\% & 71.26\% & 70.04\% & 71.48\% \\
& \textbf{HITS}        & 51.02\% & 12.61\% & 10.01\% & 27.31\% & 58.93\% & 67.68\% & 61.83\% & 67.95\% & 74.60\% & 78.22\% \\
& \textbf{TELPA}       & 64.11\% & 45.72\% & 18.94\% & 41.67\% & 71.45\% & 78.32\% & 74.88\% & 80.14\% & \underline{82.96\%} & 82.94\% \\
& \textbf{RATester}    & 67.11\% & \underline{50.83\%} & \underline{22.37\%} & \underline{45.94\%} & \underline{75.36\%} & \underline{80.45\%} & \underline{78.12\%} & \underline{83.27\%} & 81.14\% & \underline{86.55\%} \\
& \textbf{CATGen}      & \textbf{91.83\%} & \textbf{90.20\%} & \textbf{82.73\%} & \textbf{84.26\%} & \textbf{92.18\%} & \textbf{95.61\%} & \textbf{95.33\%} & \textbf{96.28\%} & \textbf{94.91\%} & \textbf{94.97\%} \\
\midrule

\multirow{7}{*}{\textbf{CovL}}
& \textbf{EvoSuite}    & 39.55\% & N/A & N/A & N/A & N/A & N/A & N/A & N/A & N/A & N/A \\
& \textbf{ChatTester}  & 30.76\% & 17.19\% & 2.51\% & 13.92\% & 40.10\% & 48.20\% & 36.76\% & 43.27\% & 36.93\% & 37.97\% \\
& \textbf{ChatUniTest} & 29.77\% & 24.75\% & 3.82\% & 12.99\% & 35.51\% & 45.92\% & 39.50\% & 46.91\% & 29.19\% & 29.35\% \\
& \textbf{HITS}        & 36.10\% & 32.89\% & 5.72\% & 18.75\% & 42.24\% & 54.47\% & 43.89\% & 46.67\% & 36.35\% & 43.95\% \\
& \textbf{TELPA}       & 45.99\% & 41.20\% & 12.35\% & 27.64\% & \underline{53.83\%} & 60.48\% & 57.31\% & 58.94\% & 49.18\% & 53.02\% \\
& \textbf{RATester}    & \underline{48.43\%} & \underline{45.73\%} & \underline{15.48\%} & \underline{30.52\%} & 49.36\% & \underline{62.47\%} & \underline{59.81\%} & \underline{63.87\%} & \underline{52.96\%} & \underline{55.68\%} \\
& \textbf{CATGen}      & \textbf{70.10\%} & \textbf{69.14\%} & \textbf{57.02\%} & \textbf{63.60\%} & \textbf{71.05\%} & \textbf{75.79\%} & \textbf{71.68\%} & \textbf{75.49\%} & \textbf{73.15\%} & \textbf{74.01\%} \\
\midrule

\multirow{7}{*}{\textbf{CovB}}
& \textbf{EvoSuite}    & 34.56\% & N/A & N/A & N/A & N/A & N/A & N/A & N/A & N/A & N/A \\
& \textbf{ChatTester}  & 29.75\% & 10.20\% & 2.52\% & 13.16\% & 45.67\% & 49.60\% & 32.00\% & 47.88\% & 32.18\% & 34.58\% \\
& \textbf{ChatUniTest} & 26.74\% & 14.92\% & 3.25\% & 10.20\% & 31.51\% & 46.78\% & 31.47\% & 49.14\% & 27.49\% & 25.93\% \\
& \textbf{HITS}        & 36.33\% & 16.61\% & 5.12\% & 12.27\% & 50.51\% & 56.45\% & 48.02\% & 57.44\% & 39.48\% & 41.04\% \\
& \textbf{TELPA}       & 44.32\% & 28.45\% & 10.32\% & \underline{25.78\%} & 56.14\% & \underline{62.83\%} & 55.81\% & 60.92\% & 48.36\% & 50.27\% \\
& \textbf{RATester}    & \underline{46.33\%} & \underline{31.67\%} & \underline{12.74\%} & 24.85\% & \underline{59.32\%} & 59.54\% & \underline{58.69\%} & \underline{65.12\%} & \underline{51.44\%} & \underline{53.61\%} \\
& \textbf{CATGen}      & \textbf{63.92\%} & \textbf{62.13\%} & \textbf{47.56\%} & \textbf{52.26\%} & \textbf{66.54\%} & \textbf{73.73\%} & \textbf{63.44\%} & \textbf{70.27\%} & \textbf{69.32\%} & \textbf{70.06\%} \\
\midrule

\multirow{7}{*}{\textbf{PR}}
& \textbf{EvoSuite}    & \underline{44.21\%} & N/A & N/A & N/A & N/A & N/A & N/A & N/A & N/A & N/A \\
& \textbf{ChatTester}  & 35.85\% & 29.27\% & 2.27\% & 16.88\% & \underline{46.42\%} & 51.57\% & 39.19\% & 52.93\% & 40.32\% & 43.78\% \\
& \textbf{ChatUniTest} & 30.24\% & 14.22\% & 3.41\% & 11.01\% & 33.54\% & 45.96\% & 31.01\% & 52.72\% & 36.96\% & 43.33\% \\
& \textbf{HITS}        & 32.93\% & 11.67\% & 3.01\% & 16.50\% & 35.17\% & 46.55\% & 48.74\% & 51.53\% & 38.12\% & 45.07\% \\
& \textbf{TELPA}       & 39.91\% & 25.48\% & \underline{12.92\%} & 22.87\% & 41.93\% & 52.41\% & 49.88\% & \underline{56.32\%} & 47.05\% & 50.36\% \\
& \textbf{RATester}    & 41.71\% & \underline{29.76\%} & 10.21\% & \underline{25.34\%} & 42.25\% & \underline{55.93\%} & \underline{54.32\%} & 54.12\% & \underline{49.87\%} & \underline{53.61\%} \\
& \textbf{CATGen}      & \textbf{54.63\%} & \textbf{47.00\%} & \textbf{34.71\%} & \textbf{37.62\%} & \textbf{50.54\%} & \textbf{67.77\%} & \textbf{64.53\%} & \textbf{69.36\%} & \textbf{58.34\%} & \textbf{61.77\%} \\
\midrule
\end{tabular}}

\vspace{-8pt}
\end{table}

%% file: Tables/rq2.tex
\begin{table}[]

\caption{Effectiveness comparison between \tech{} and its variants.}
\label{tab:rq2}
\vspace{-10pt}
\small
\renewcommand{\arraystretch}{1}
\begin{tabular}{l|rrrr}
\toprule
\textbf{Method}                  & \textbf{CSR}     & \textbf{CovL}    & \textbf{CovB}    & \textbf{PR}         \\ 
\midrule
\textbf{CATGen}                  & \textbf{95.61\%} & \textbf{75.79\%} & \textbf{73.73\%} & \textbf{57.77\%}    \\
\textbf{$\Delta$w/o skeleton}    & -3.04\%          & -16.37\%         & -10.28\%         & -8.86\%             \\
\textbf{$\Delta$w/o repair}      & -16.79\%         & -8.95\%          & -6.22\%          & -8.76\%             \\
\textbf{$\Delta$w/o enhancement} & -1.96\%          & -10.51\%         & -9.51\%          & -0.13\%             \\
\textbf{$\Delta$w/o all}         & -38.88\%         & -30.63\%         & -24.00\%         & -15.98\%            \\
\bottomrule
\end{tabular}
\vspace{-10pt}
\end{table}

%% file: Tables/rq3.tex
\begin{table}[]
\centering
\caption{Efficiency comparison between \tool{} and baselines.}
\label{tab:rq3}
\vspace{-10pt}
\small
\setlength{\tabcolsep}{3pt}
\renewcommand{\arraystretch}{1.0}
\begin{tabular}{l|rrrrrr}
\toprule
\textbf{Method}             & \textbf{Gen.Time}  & \textbf{Gen.Tokens}  & \textbf{Post.Time}  & \textbf{Post.Tokens}  & \textbf{Total.Time}  & \textbf{Total.Tokens}  \\ \midrule
\textbf{EvoSuite}           & 10,980s             & 0k                   & 0s                  & 0k                   &10,980s               & 0k                     \\
\textbf{ChatTester}         & 2,805s             & 432k                 & 1,666s              & 459k                  & 4,471s               & 891k                   \\
\textbf{ChatUniTest}        & \textbf{1,601s}    & 234k                 & 2,166s              & 857k                  & 3,768s               & 1,091k                 \\
\textbf{HITS}               & 2,972s             & 459k                 & 2,951s              & 799k                  & 5,924s               & 1,258k                 \\
\textbf{TELPA}              & 4,386s             & 612k                 & 0s                  & 0k                    & 4,386s               & 612k                   \\
\textbf{RATester}           & 5,217s             & 735k                 & 0s                  & 0k                    & 5,217s               & 735k                   \\
\textbf{CATGen}             & 1,759s             & \textbf{203k}        & \textbf{77s}        & \textbf{0k}           & \textbf{1,836s}      & \textbf{203k}          \\
\bottomrule
\end{tabular}
\vspace{-10pt}
\end{table}

%% file: 5_oss_evaluation.tex
\section{Open-Source Software Evaluation}
\label{sec:oss}
To assess the generalizability of \tech{} beyond industrial codebases, we additionally evaluate it on the widely used open-source benchmark \textbf{Defects4J}~\cite{just2014defects4j}. Following prior studies~\cite{yang2025clarifying}, we select four representative projects, \textit{Chart}, \textit{Lang}, \textit{Time}, and \textit{Math}, which span diverse domains such as chart rendering, language utilities, date-time processing, and numerical computation.

We reuse the same focal-method evaluation protocol, comparison techniques, metrics, and experimental configuration as in the industrial evaluation.
To ensure a controlled comparison in the OSS setting, we instantiate CATGen and all LLM-based baselines with a single underlying model, \textbf{Qwen2.5-Coder-32B-Instruct (QC-32B)}, for all LLM invocations.
All other factors, including prompts, inference settings, compilation and execution pipelines, and cost accounting, are kept identical. 
This setup isolates the effect of the benchmark itself, ensuring that any observed differences arise from dataset characteristics rather than changes in experimental configuration.

To further evaluate fault-detection capability, we report \textbf{Mutation Score (MS)}---the ratio of killed mutants to total mutants---which directly evaluates fault-detection effectiveness beyond PR.
MS is reported only on the open-source benchmark; it is not reported for the industrial setting due to the proprietary nature of the dataset.

\input{Tables/rq4}

\textbf{Finding V: CATGen remains the most effective on open-source benchmarks compared to LLM-based approaches.} 
As shown in Table~\ref{tab:rq4}, on Defects4J, existing approaches already achieve strong results in terms of both compilation success and structural coverage, indicating that current methods can generate usable unit tests for open-source projects with relatively stable environments.
Under this strong baseline performance, CATGen still achieves the best overall effectiveness among LLM-based methods. 
In particular, CATGen attains the highest compilation success rate (92.40\%) and pass rate (79.65\%) among all LLM-based techniques, and remains only slightly below EvoSuite in compilation success (98.17\%). 
This is consistent with practical experience that Defects4J-style projects, which mainly contain computation-oriented logic with limited framework dependencies, are particularly favorable to search-based tools such as EvoSuite.
At the same time, CATGen substantially improves test adequacy: it achieves the highest line coverage (80.64\%) and branch coverage (75.58\%) across all compared methods, surpassing both traditional search-based generation (EvoSuite: 62.44\% / 53.08\%) and the strongest LLM-based baselines (e.g., RATester: 72.85\% / 66.32\%). 
These results indicate that CATGen improves coverage while preserving a high level of executability.

From our experience, the open-source results align with what we observed in the industrial evaluation: stabilizing the test class skeleton reduces early compilation friction, and lightweight post-processing helps recover common issues without relying on iterative LLM repair. In open-source projects where the environment is generally cleaner, these mechanisms appear to function as “reliability amplifiers”—they do not replace strong baseline capabilities, but make the generated tests more consistently executable and more thorough in structural exploration.

\begin{table}[t]
\centering
\setlength{\abovecaptionskip}{2pt}
\setlength{\belowcaptionskip}{-5pt}
\caption{Mutation Score (MS) on Defects4J.}
\label{tab:ms}
\begin{tabular}{lccccccc}
\toprule
\textbf{Project} &
\textbf{EvoSuite} & \textbf{ChatTester} & \textbf{ChatUniTest} & \textbf{HITS} & \textbf{TELPA} & \textbf{RATester} & \textbf{CATGen} \\
\midrule
Chart & 58.2\% & 50.1\% & 48.6\% & 53.4\% & 57.8\% & \underline{60.2\%} & \textbf{66.9\%} \\
Math  & 63.5\% & 54.6\% & 52.1\% & 57.2\% & 61.3\% & \underline{64.1\%} & \textbf{71.2\%} \\
Lang  & 65.8\% & 57.9\% & 55.4\% & 60.3\% & 64.7\% & \underline{66.5\%} & \textbf{73.6\%} \\
Time  & 57.6\% & 48.8\% & 47.2\% & 52.1\% & 55.9\% & \underline{58.7\%} & \textbf{67.4\%} \\
\midrule
\textbf{Avg} & 61.3\% & 52.8\% & 50.8\% & 55.9\% & 59.7\% & \underline{62.4\%} & \textbf{69.8\%} \\
\bottomrule
\end{tabular}
\end{table}

\textbf{Finding VI: CATGen improves fault-detection capability as measured by mutation score.}
To further assess whether the generated tests are effective in detecting faults beyond executability, we evaluate the mutation score (MS) on Defects4J. As shown in Table~\ref{tab:ms}, CATGen consistently achieves the highest mutation score across all projects, outperforming the strongest baseline (RATester) by 7.4 percentage points on average. This indicates that the improvements in compilation success and structural coverage also translate into stronger fault-detection capability.

Interestingly, although some baselines achieve relatively high line or branch coverage, their mutation scores remain comparatively lower, suggesting that high structural coverage does not necessarily imply effective fault detection. In contrast, CATGen not only generates executable tests but also produces tests that are more effective in killing injected mutants. These results highlight the importance of complementing traditional metrics such as PR and coverage with mutation-based evaluation, and confirm that CATGen improves both executability and fault-detection effectiveness.

%% file: Tables/rq4.tex
\begin{table*}[]
\centering
\setlength{\abovecaptionskip}{2pt}
\setlength{\belowcaptionskip}{-5pt}
\renewcommand{\arraystretch}{1.12}
\setlength{\tabcolsep}{5pt}
\footnotesize
\caption{Effectiveness on Defects4J.}
\label{tab:rq4}
\resizebox{\textwidth}{!}{
\begin{tabular}{llccccccc}
\toprule
\textbf{Metric} & \textbf{Project} &
\textbf{EvoSuite} & \textbf{ChatTester} & \textbf{ChatUniTest} & \textbf{HITS} & \textbf{TELPA} & \textbf{RATester} & \textbf{CATGen} \\
\midrule

\multirow{4}{*}{\textbf{CovL}} 
& \textbf{Chart} & 48.31\% & 54.24\% & 59.10\% & 62.14\% & 64.50\% & \underline{67.53\%} & \textbf{75.71\%} \\
& \textbf{Math}  & 60.21\% & 47.89\% & 51.81\% & 54.42\% & 64.56\% & \underline{67.13\%} & \textbf{73.32\%} \\
& \textbf{Lang}  & 71.32\% & 80.47\% & 79.12\% & \underline{83.52\%} & 81.21\% & 80.42\% & \textbf{91.32\%} \\
& \textbf{Time}  & 69.91\% & 71.33\% & 70.21\% & 75.31\% & 76.24\% & \underline{76.31\%} & \textbf{82.21\%} \\
\midrule

\multirow{4}{*}{\textbf{CovB}} 
& \textbf{Chart} & 43.51\% & 48.14\% & 51.28\% & 60.47\% & 60.74\% & \underline{62.43\%} & \textbf{72.89\%} \\
& \textbf{Math}  & 45.17\% & 38.54\% & 40.78\% & 48.21\% & \underline{61.54\%} & 54.32\% & \textbf{64.74\%} \\
& \textbf{Lang}  & 62.89\% & 74.21\% & 76.32\% & 77.21\% & \underline{80.21\%} & 78.21\% & \textbf{84.25\%} \\
& \textbf{Time}  & 60.74\% & 65.33\% & 67.32\% & 69.24\% & \underline{75.87\%} & 70.32\% & \textbf{80.45\%} \\
\midrule

\multirow{4}{*}{\textbf{CSR}} 
& \textbf{Chart} & \textbf{98.21\%} & 64.32\% & 71.21\% & 75.41\% & 76.54\% & 80.12\% & \underline{90.45\%} \\
& \textbf{Math}  & \textbf{97.24\%} & 59.21\% & 65.78\% & 67.89\% & 70.12\% & 77.32\% & \underline{91.57\%} \\
& \textbf{Lang}  & \textbf{100.00\%} & 71.54\% & 76.20\% & 82.65\% & 83.41\% & 82.12\% & \underline{94.45\%} \\
& \textbf{Time}  & \textbf{97.21\%} & 69.54\% & 70.24\% & 77.54\% & 75.69\% & 80.23\% & \underline{93.14\%} \\
\midrule

\multirow{4}{*}{\textbf{PR}} 
& \textbf{Chart} & \textbf{89.45\%} & 41.23\% & 59.10\% & 62.14\% & 64.50\% & 60.34\% & \underline{78.12\%} \\
& \textbf{Math}  & \textbf{90.21\%} & 44.89\% & 51.81\% & 54.42\% & 64.56\% & 65.21\% & \underline{75.78\%} \\
& \textbf{Lang}  & \textbf{89.12\%} & 58.54\% & 65.32\% & 69.54\% & 74.56\% & 75.32\% & \underline{83.24\%} \\
& \textbf{Time}  & \textbf{90.42\%} & 60.87\% & 59.45\% & 70.24\% & 72.21\% & 72.56\% & \underline{81.45\%} \\
\bottomrule
\end{tabular}}
\end{table*}

%% file: 6_discussion.tex
\section{Discussion}
\label{sec:discussion}

\textbf{I. Case Study and Practical Insights}.
\begin{figure}[t]
    \centering
    \includegraphics[width=0.88\linewidth]{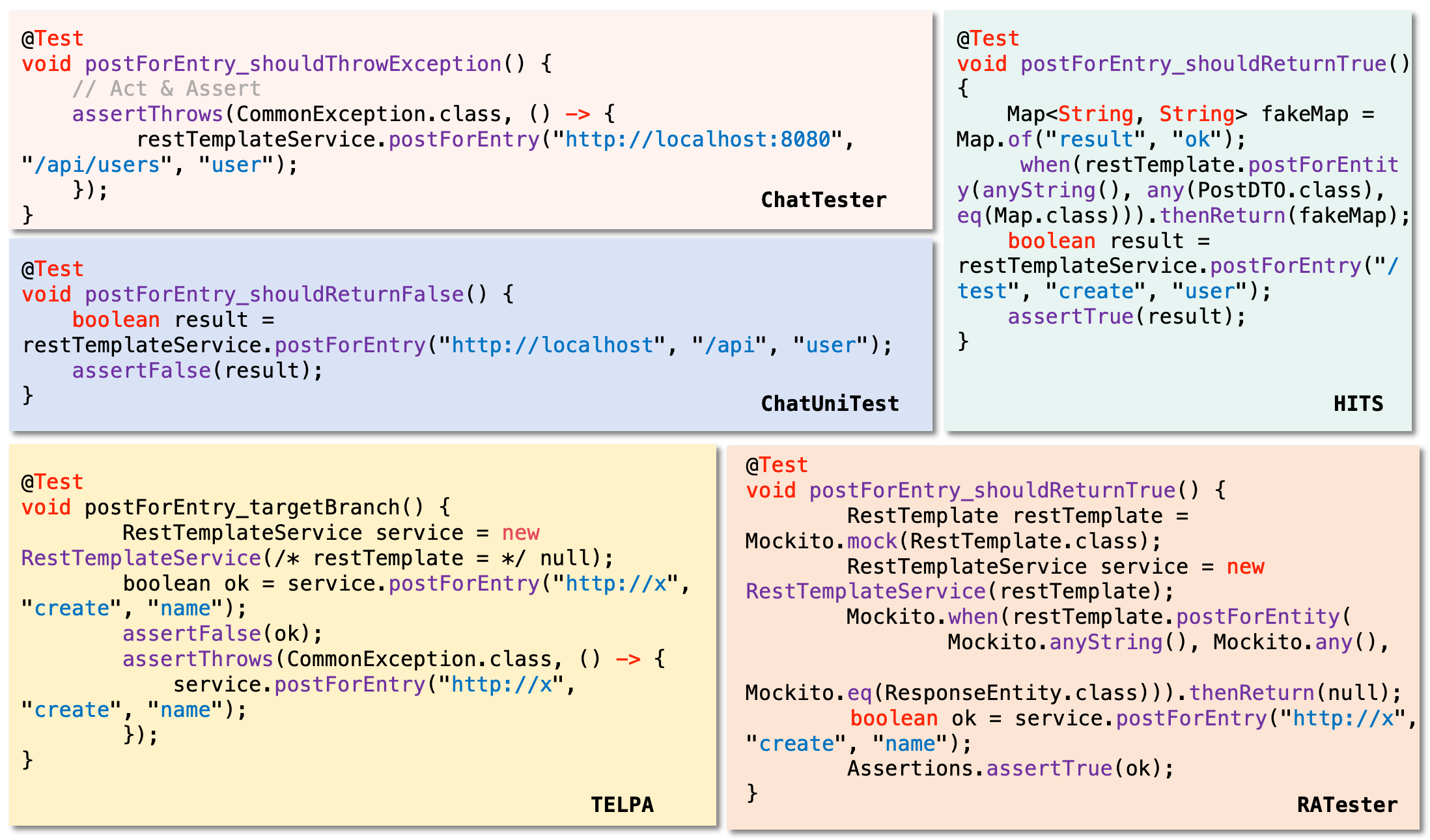}
    \caption{Unit test generated by baselines}
    \label{fig:baselines}
\end{figure}
We now use the case in Figure~\ref{fig:test} (QC-32B) to qualitatively demonstrate how \tech{} operates and outperforms baselines. The corresponding focal method is shown in Figure~\ref{fig:focal}.
Based on the structured test class skeleton, \tech{} demonstrates clear advantages in test effectiveness over existing baselines. 
It generates test methods that accurately cover both nominal and exceptional execution paths.
In the successful case, \tech{} configures \textit{restTemplate.postForEntity()} via \textit{when/thenReturn} to return a properly typed \textit{ResponseEntity} containing expected data (lines 4), thereby exercising the focal method’s main control flow. 
In the exceptional case, it simulates a network failure by injecting \textit{when().thenThrow(new RestClientException("Error"))} (lines 11), ensuring that the focal method properly handles the exception and rethrows it as a \textit{CommonException} (lines 13-14). 
Additionally, \tech{}'s program analysis-based enhancement component identifies unhandled edge conditions, such as \textit{null} values, which are often overlooked by LLMs, and generates dedicated tests to cover these cases explicitly. 
By integrating these three test methods into a cohesive and well-structured test class, CATGen achieves comprehensive functional coverage of the focal method.
For the same focal method, EvoSuite generated tests that failed to compile in our setting, which reflects a recurring challenge we encountered in industrial repositories where framework wiring, external types, and required initialization must be correct before any test logic becomes executable. 

When turning to LLM-based baselines in Figure~\ref{fig:baselines}, we observed different practical tradeoffs. ChatTester and ChatUniTest tend to exercise the system under real or close-to-real runtime conditions, which makes it difficult to reproduce exception scenarios such as network failures in a controlled test environment without inducing actual faults. 
HITS does incorporate mocking, yet it can still produce mocks that do not match the intended service behavior, for example returning a raw \textit{Map} where a \textit{ResponseEntity} is required, which weakens behavioral fidelity. 
TELPA is guided by coverage feedback and often tries to reach uncovered branches quickly, but in our setting it can be sensitive to harness stability, as shown by directly constructing \textit{RestTemplateService} with a \textit{null} dependency and invoking \textit{postForEntry} without consistently configuring \textit{postForEntity} via \textit{when} and \textit{thenReturn} or \textit{thenThrow}. As a result, the generated tests may fail early due to incomplete initialization.
RATester enriches generation with repository-level symbol and usage information and does set up \textit{Mockito.mock} with a \textit{when} rule for \textit{restTemplate.postForEntity}, yet the test still hinges on scaffolding and mock semantics, such as returning \textit{null} for a call that is expected to produce a \textit{ResponseEntity}. This kind of mismatch can shift execution into a different path and reduces how faithfully the test represents the target service behavior under industrial project conventions.
These observations suggest that, for industrial projects, incorporating broader context is helpful, while stabilizing the test class skeleton and initialization remains a prerequisite for turning that context into compilable and behavior-valid tests.
Taken together, the quantitative evaluations and qualitative case analysis strongly reinforce the effectiveness of \tech{}’s multi-stage architecture in addressing the challenges of LLM-based unit test generation, particularly in industrial scenarios. 

\noindent \textbf{II. Lessons from Replacing LLM-Based Repair with Static Analysis}.
Our results show that deterministic program analysis yields more effective post-processing than probabilistic, LLM-based repair. To validate this claim, we further investigate the extent to which \tech{}’s program analysis-based post-processing can enhance existing baselines.
Specifically, we collect the baseline results before post-processing, then substitute each baseline’s original post-processing module with \tech{}’s, and re-evaluate the metrics. We use the subscript \textit{w} to denote results before repair and \textit{c} for results after applying \tech{}’s post-processing.
As in previous research questions, we use QC-32B as the underlying LLM for consistency.

\input{Tables/discussion_post}

Results presented in Table~\ref{tab:discussion} demonstrate that integrating \tech{}'s post-processing module into baselines consistently enhances performance across all evaluation metrics.
For CovL, ChatTester's performance increases from 48.20\% to 49.95\%, ChatUniTest's from 45.92\% to 50.81\%, and HITS's from 54.47\% to 59.01\%. 
In CovB, ChatTester improves from 49.60\% to 50.17\%, ChatUniTest from 46.78\% to 56.20\%, and HITS from 56.45\% to 63.70\%.
CSR rises notably: ChatTester's CSR increases from 73.03\% to 83.49\%, ChatUniTest's from 71.06\% to 80.04\%, and HITS's from 67.68\% to 85.28\%. 
Similarly, in PR, ChatTester improves from 51.57\% to 52.82\%, ChatUniTest from 45.96\% to 48.51\%, and HITS from 46.55\% to 53.35\%. 
Notably, all baselines without post-processing exhibit inferior performance, with CSR dropping below 66\% and PR decreasing below 46\%. 
This stark contrast highlights the critical importance of post-processing in unit test generation.
These results validate that deterministic program analysis outperforms heuristic strategies in post-processing, particularly for tasks requiring rigorous adherence to code correctness and execution stability.
This suggests that making repair deterministic and context-grounded is a stronger leverage point than adding more LLM iterations, because it bounds cost while systematically eliminating recurring compiler-level failure modes.

\noindent \textbf{III. Future Work}.
Currently, \tech{} primarily targets Java projects; future work will extend its capabilities to support mainstream programming languages such as Python, C++, and JavaScript, enabling broader applicability across diverse software ecosystems.  
A key focus will be optimizing model inference efficiency by refining inference workflows and post-processing strategies to enhance scalability in large-scale industrial projects, reducing both generation time and computational overhead.  
To improve test robustness, we will investigate methods for generating more diverse test cases that comprehensively cover edge cases, boundary conditions, and exception scenarios.  This includes developing advanced condition-decomposition algorithms and boundary-value analysis techniques to address under-tested scenarios systematically.

\section{Threats to Validity}
\label{subsec:threats}

\textit{External Validity}. A potential threat to external validity arises from the choice of models and the implementation of baselines. To mitigate this, we evaluate \tech{} across multiple state-of-the-art LLMs with diverse architectures, including general-purpose and code-specialized models, to ensure findings are not tied to a single model. For baselines, we adopt publicly available implementations to ensure fair and reproducible comparisons.
Our evaluation targets reproducible, pipeline-level test generation with locally hosted open-weight models; we do not report head-to-head experiments against interactive coding assistants (e.g., Claude Code and Codex-based tools), which optimize ad hoc developer interaction rather than batch generation under fixed project context and may raise data-security concerns for proprietary code under NDA settings.

\noindent
\textit{Internal Validity}. Three aspects affect internal validity. First, our coverage metrics rely on Jacoco for runtime instrumentation, which may miss certain execution paths or misreport coverage in edge cases. While Jacoco is standard in Java testing, this limitation is acknowledged. 
Second, the non-deterministic nature of LLMs may introduce variability in generated tests. To address this, we conduct multiple independent runs for each configuration and apply statistical significance testing to ensure observed improvements are consistent.
Third, when contrasting \tech{} with baselines in Section~\ref{sec:industrial}, we compare outcomes on identical focal methods using paired Wilcoxon signed-rank tests~\cite{woolson2007wilcoxon}, reporting $p$-values together with effect sizes ($r$) and the accompanying changes in compilation success rate, line coverage, and branch coverage.

\noindent
\textit{Construct Validity}. We treat success as compilable, executable tests under project constraints, not metric inflation alone. The industrial benchmark stresses deployment failure modes and may favor compilability-focused workflows; we therefore complement it with Defects4J and mutation-score analysis. Industrial focal methods were not publicly available during LLM training, and generic repair patterns also transferred to C pilots, but all reported experiments are Java-based.
Static context extraction uses IntelliJ PSI in our implementation but depends only on AST-level structural signals; any extractor providing equivalent facts can substitute for PSI.

\vspace{-5pt}

%% file: Tables/discussion_post.tex
\begin{table}[t]
\caption{Effectiveness comparison of different post-processing strategies.}
\label{tab:discussion}
\vspace{-10pt}
\footnotesize
\renewcommand{\arraystretch}{1}  

\begin{tabular}{l|cccc}
\toprule
\textbf{Method}                          & \textbf{CovL}    & \textbf{CovB}    & \textbf{CSR}     & \textbf{PR}       \\
\midrule
\textbf{ChatTester}                      & 48.20\%          & 49.60\%          & 73.03\%          & 51.57\%           \\
\textbf{ChatTester\textsubscript{w}}     & 43.93\%          & 43.57\%          & 63.36\%          & 44.32\%           \\
\textbf{ChatTester\textsubscript{c}}     & \textbf{49.95\%} & \textbf{50.17\%} & \textbf{83.49\%} & \textbf{52.82\%}  \\
\midrule
\textbf{ChatUniTest}                     & 45.92\%          & 46.78\%          & 71.06\%          & 45.96\%           \\
\textbf{ChatUniTest\textsubscript{w}}    & 38.53\%          & 40.05\%          & 59.05\%          & 38.33\%           \\
\textbf{ChatUniTest\textsubscript{c}}    & \textbf{50.81\%} & \textbf{56.20\%} & \textbf{80.04\%} & \textbf{48.51\%}  \\
\midrule
\textbf{HITS}                            & 54.47\%          & 56.45\%          & 67.68\%          & 46.55\%           \\
\textbf{HITS\textsubscript{w}}           & 48.18\%          & 50.19\%          & 65.30\%          & 41.75\%           \\
\textbf{HITS\textsubscript{c}}           & \textbf{59.01\%} & \textbf{63.70\%} & \textbf{85.28\%} & \textbf{53.35\%}  \\
\bottomrule
\end{tabular}

\begin{tablenotes}
\centering  
\small
\item Note: w = without post-processing; c = with CATGen's post-processing component.
\end{tablenotes}
\vspace{-8pt} 
\end{table}

%% file: 7_conclusion.tex
\section{Conclusion}
\label{sec:conclusion}
LLMs have recently shown promise for automated unit test generation, yet our industrial deployments reveal a persistent gap between encouraging research results and practical usability.
In real-world projects, generated tests must not only achieve coverage but also compile and execute under complex project-level constraints.
In our experience, we observed that insufficient code context, fragile scaffolding, and costly repair loops frequently dominate the workflow, limiting the benefits of existing prompt-driven approaches.
To tackle this, we introduce a context-aware workflow developed through repeated industrial failures. 
Rather than relying on LLMs to infer missing dependencies, CATGen explicitly retrieves project context, constructs deterministic test-class skeletons, and applies analysis-driven post-processing to ensure compilation robustness.
Across both industrial and open-source settings, CATGen consistently generates more compilable and effective tests while significantly reducing generation cost compared to existing LLM-based approaches.
Our insights encourage future work to treat LLMs as components within engineered systems rather than standalone solutions, and to prioritize robustness and deployability when bringing AI-assisted testing techniques into real-world development environments.


\section*{Data Availability}
\label{sec:data_availability}

Due to confidentiality agreements with our industrial partner, the full source code and proprietary benchmark datasets used in this work cannot be publicly released.
However, to maximize reproducibility and transparency under these constraints, we maintain a public replication repository that documents evaluated LLM checkpoints and inference settings; the scope of static context extraction and its portability across representative IDE parser APIs; deterministic, framework-aware test skeleton construction; the compilation-repair strategies in Section~\ref{sec:3.4} with worked examples and ordering rationale; consolidation rules for merging LLM-generated fragments with analysis-driven enhancements into one test class; and supporting Java parsing utilities together with auxiliary preprocessing and results-processing scripts released with the package.
We additionally provide anonymized prompts, representative focal-method examples, and intermediate results aligned with these materials.\footnote{\url{https://github.com/CATGen-repository/CATGen}}